\documentclass[]{aa}

\usepackage{graphicx}
\usepackage{natbib}
\usepackage{txfonts}
\usepackage{upgreek}

\begin{document} 

\title{Colors of an Earth-like exoplanet}
\subtitle{Temporal flux and polarization signals of the Earth}

\author{A. Groot
          \inst{1},
        L. Rossi
          \inst{2},
        V.J.H. Trees
          \inst{1}
          \fnmsep\thanks{Now at Royal Netherlands Meteorological Institute (KNMI), De Bilt, The Netherlands.},
        J.C.Y. Cheung
          \inst{1}
          \and
        D.M. Stam
          \inst{1}
        }

\institute{Faculty of Aerospace Engineering, 
           Delft University of Technology, Delft, The Netherlands \\
           \email{d.m.stam@tudelft.nl}
        \and
            LATMOS/IPSL, UVSQ Université Paris-Saclay, Sorbonne Université, CNRS, Guyancourt, France
           }
           
\date{Received January 24, 2020; accepted June 29, 2020}

\abstract
{Understanding the total flux and polarization signals of Earth-like
planets and their spectral and temporal variability is essential for 
the future characterization of such exoplanets.}
{We provide computed total ($F$) and linearly ($Q$ and $U$) and circularly
($V$) polarized fluxes, and the degree of polarization $P$ of sunlight
that is reflected by a model Earth, to be used for instrument designs,
optimizing observational strategies, and/or developing retrieval algorithms.} 
{We modeled a realistic Earth-like planet using one year of daily 
Earth-observation data: cloud parameters (distribution, optical 
thickness, top pressure, and particle effective radius), and surface
parameters (distribution, surface type, and albedo). The Stokes vector of
the disk-averaged reflected sunlight was computed for phase angles 
$\alpha$ from 0$^\circ$ to 180$^\circ$, and for wavelengths $\lambda$ 
from 350 to 865~nm.}
{The total flux $F$ is one order of magnitude higher than the polarized flux
$Q$, and $Q$ is two and four orders of magnitude higher than $U$ and
$V$, respectively. Without clouds, the peak-to-peak daily variations
due to the planetary rotation increase with increasing $\lambda$ for 
$F$, $Q$, and $P$, while they decrease for $U$ and $V$. Clouds modify
but do not completely suppress the variations that are due to rotating
surface features. With clouds, the variation in $F$ increases with
increasing $\lambda$, while in $Q$, it decreases with increasing
$\lambda$, except at the largest phase angles. In earlier work, it was
shown that with oceans, $Q$ changes color from blue through 
white to red. The $\alpha$ where the color changes increases with 
increasing cloud coverage. Here, we show that this unique color 
change in $Q$ also occurs when the oceans are partly replaced by 
continents, with or without clouds. The degree of polarization
$P$ shows a similar color change. 
Our computed fluxes and degree of polarization will be made publicly
available.
}
{}


\authorrunning{A. Groot et al.}
\maketitle

\section{Introduction}
\label{Sect_Introduction}

Since the discovery of 51 Pegasi b \citep{mayor1995jupiter}, several
thousand exoplanets have been discovered and 
confirmed\footnote{http://exoplanetarchive.ipac.caltech.edu} 
including various Earth-sized planets in the habitable zones around 
their parent stars \citep{choukeplercatalog}. A recent addition to the
latter selection is TOI-700b, an Earth-sized planet in the habitable 
zone of red dwarf star TOI-700, as discovered by 
\citet{2020TOI700Discovery} and confirmed with Spitzer
observations by \citet{2020TOI700confirmationSpitzer}.

The planetary physical properties that can be derived from measurements 
with the two most successful exoplanet detection techniques, that is, with 
the radial velocity method and the transit method, are the planet 
radius (if the planet transits its star), mass (a lower limit if the
planet does not transit its star), and the orbital period and distance,
eccentricity, and inclination (if the planet transits its star)
\citep[see][and references therein]{2018exha.book.....P}. 
The upper atmospheres (above optically thick clouds) of giant 
exoplanets and those of smaller planets around M stars can be probed 
using spectroscopy during primary and/or secondary transits 
\citep{2010Natur.468..669B,swain2009water,swain2008presence,tinetti2007water,NakajimaandTanaka1983}. 
It appears to be virtually impossible to characterize the (lower)
atmospheres and surfaces of small, Earth-like planets in the habitable
zones of solar-type stars \citep[][]{B_tr_mieux_2014,Misra_2014,kaltenegger2009transits}, 
in particular because the light of the parent star is refracted while
traveling through the lower atmosphere of its planet and emerges forever 
out of reach of terrestrial telescopes \citep[][]{2012ApJ...755..103G}.
The (lower) atmosphere and surface of a planet are crucial for 
determining the habitability of a planet, as they hold information about 
cloud composition, trace gases in disequilibrium and probably most
importantly, liquid surface water \citep[see, e.g.,][and references therein]{2018AsBio..18..663S,2007AsBio...7..252K,2007AsBio...7..222K}.
For such a characterization of terrestrial-type planets, direct 
observations of the thermal radiation that they emit or of the light 
of their parent star that they reflect are required. The numerical 
results that we present in this paper concern the reflected starlight.
Because of the huge distances involved, any measured reflected starlight 
pertains to the (illuminated and visible part of the) planetary disk.
It therefore is a disk-integrated signal. 

\citet{ford2001characterization} showed with numerical simulations
that the daily 
rotation of the Earth induces repetitive patterns in 
the disk-integrated flux that is reflected by the Earth
observed from afar because of the continents and oceans and the clouds.
\citet{stam2008spectropolarimetric} showed how not only the 
total flux of reflected sunlight, but also the polarized fluxes
and the degree of polarization of this light vary as
an Earth-like planet rotates and orbits its star. 
Recently, \citet{TreesandStam2019} showed the change in the color of the 
polarized fluxes and degree of polarization throughout the 
orbit when an ocean is present. 
Polarimetry indeed appears to be a powerful tool for the detection 
and characterization of exoplanets. 
In particular, when it is integrated over the stellar disk, 
direct starlight is virtually 
unpolarized \citep{1987Kemp}, while starlight that has been scattered in
a planetary atmosphere and/or has been reflected by a planetary 
surface is usually polarized. 
Polarimetry thus enhances the contrast between a star and an orbiting 
exoplanet 
(by three to five orders of magnitude, according to \citet[][]{Keller2010}), 
allows the direct confirmation of the nature of the 
substellar companion, and the state of polarization of the reflected
light holds information about the composition and structure of the
planetary atmosphere and surface (if present) 
\citep[see][for details and references]{2015psps.book..439W}.

A recent polarimetric detection in reflected light
of a substellar companion
that is possibly surrounded by a disk was announced 
by \citet{2018A&A...616A..79G}.
The recent tentative polarimetric detection of banded,
that is, Jupiter-like, cloud structures on Luhman 16A by 
\cite{2020ApJ...894...42M}
supports the application of polarimetry for exoplanet 
characterization (in this case in the thermally emitted signal
of this binary brown dwarf system component).

In this paper, we present the computed phase curves of the 
total and polarized fluxes and the degree of polarization of 
starlight that is reflected by an Earth-like exoplanet 
with a spatially and temporally changing cloud and surface cover, 
all based on daily data from Earth-observation satellite
instruments.
The temporal resolution of our computations is 15~minutes, and it covers
2011 completely. The phase angle ranges from 0$^\circ$ to 180$^\circ$.
We did not include gaseous absorption,
and instead concentrated on the broadband continuum fluxes at 350,
443, 550, 670, and 865~nm. 

Our computed phase curves are publicly available and can be used for the
design of future instruments 
for the detection and characterization 
of terrestrial-type exoplanets.
The polarization state of reflected starlight also needs to be included 
for the
design of instruments that aim at measuring only total fluxes,
because the response of optical elements such as mirrors and gratings 
usually depends on the polarization state of the incident light. 
The curves can also be used for the development and testing of retrieval algorithms
\citep[see, e.g.,][]{2020arXiv200403941A,2019ApJ...882L...1F,Berdyugina_2019,2011ApJ...739L..62K}, and 
for the optimization of observational strategies.
A number of references describing the use of terrestrial remote-sensing
satellite measurements of the total flux of sunlight reflected 
by Earth for the testing of retrieval algorithms are provided by
\citet{2018AJ....156...26J}.  
Fluxes that have been measured in this way, however, usually do not cover the
phase angle range and/or do not have the temporal resolution and/or 
do not cover the spectral range 
to fully assess (future) measurements of exoplanet fluxes.
Earth-shine measurements \citep[see][and references to these papers]{2006ApJ...644..551T,2002ApJ...574..430W}, 
where the sunlight that is reflected by 
the Earth is detected on the nightside of the moon, are also 
limited by the phase angle range and the lack of knowledge of the
reflection properties of the lunar surface.
Computed reflected fluxes do not suffer from these limitations,
and the user has the ability to add 
specific noise and/or instrumental effects.

As far as we know, the polarized fluxes
of the (distant) Earth have never been measured directly. 
Linearly polarized fluxes of local regions on Earth have been
measured by the POLDER-3/PARASOL Low-Earth-Orbit 
satellite instrument \citep[]{2007ApOpt..46.5435F}
and its predecessors POLDER-1 and POLDER-2 \citep[]{1994ITGRS..32..598D}.
\citet[]{2019JQSRT.224..474D} presented an excellent
review of the characterization of aerosol and surface properties
from such local data. Attempts have been made to translate
local POLDER-3/PARASOL data into a polarized signal that 
simulates the disk-integrated signal of the Earth 
\citep[]{2019RAA....19..117W}, but especially because 
polarization signals depend strongly on the illumination
and viewing geometries \citep[see, e.g.,][]{Hansen1974}, 
the results of these attempts remain very limited.
First spectropolarimetric measurements of Earth-shine have been presented
by \citet{2014A&A...562L...5M}, \citet{2013A&A...556A.117B}, 
and \citet{2012Natur.483...64S}, 
and a more in-depth analysis of the latter data
can be found in \citet[][]{Sterzik2018} and
\citet[][]{2017A&A...605A...2E}. In these measurements, the 
reflection properties of the lunar surface,
in particular, the reflection of polarized incident Earth-shine
by the lunar surface, indeed remains a source of uncertainty. 

A few instrument developments have aimed at measuring total 
and polarized 
fluxes of sunlight that is reflected by the Earth from afar, such as
the Lunar Observatory of Unresolved Polarimetry of Earth (LOUPE)
\citep{LOUPE2016,2012P&SS...74..202K} that is scheduled to observe the Earth
from the surface of the moon (or, e.g., a geostationary satellite, which
would mean that we would not be able to monitor the daily rotation of the 
Earth, but which would ensure continuous access to power). 
Examples of future telescope instruments that are designed to 
measure polarization of exoplanets are EPOL, the imaging polarimeter 
of the Exoplanet Imaging Camera and Spectrograph (EPICS) for the 
European Extremely Large Telescope (E-ELT) \citep{Kasper2010,Keller2010}, 
and POLLUX, a UV polarimeter, that is envisioned for the NASA 
Large UV Optical Infrared Surveyor (LUVOIR) space telescope concept 
\citep{Bouret2018}.

A number of publications has focused on the simulation of reflected-light
signals of Earth-like exoplanets (see the various references in this article),
but did not have the temporal resolution of our results. 
Two other important factors in our model are the reflection by the 
wind-ruffled ocean surface \citep[as described by][]{TreesandStam2019}
and the realistic cloud properties. 
Simulations of light reflected by exo-oceans have been presented by  
\citet{zugger2010light} and \citet{williams2008detecting}.
They used models without full polarization, with simplified cloud
models, and without cloud and surface variability.

The outline of this paper is as follows. 
In Sect.~\ref{Sect_Numerical_Algorithms} we describe the numerical 
algorithm that we used to compute the fluxes and polarization signals of 
our model planets, where we start with our definitions and end with 
a concise description of the radiative transfer algorithm.
In Sect.~\ref{Sect_Exoplanetmodels}, we describe the parameters of the
model planet and the databases from which we have obtained these parameters.
In Sect.~\ref{Sect_Results}, we present the computed total and
polarized fluxes and degree of polarization for cloud-free and cloudy
model planets. Finally, in Sect.~\ref{Sect_Discussion}, we summarize and 
discuss our results and describe possible improvements of the numerical 
simulations and the relation with (future) observations.

\section{Numerical algorithms}
\label{Sect_Numerical_Algorithms}

\subsection{Definitions of light and polarization}
\label{Sect_Definitions}

We describe the starlight that is incident on a planet and the 
starlight that is reflected by the planet by the Stokes vector 
${\bf F}$ \citep[see, e.g.,][]{Hovenier2004,Hansen1974},
\begin{equation}
   \mathbf{F} = \left[ \begin{array}{c}
                         F \\ Q \\ U \\ V
                         \end{array} \right],
\label{eq:def-stokes-vector}
\end{equation}
where $F$ is the total flux, $Q$ and $U$ are the linearly 
polarized fluxes, and $V$ is the circularly polarized flux 
(all in W~m$^{-2}$, or W~m$^{-3}$ when defined per wavelength). 
Fluxes $Q$ and $U$ are defined with respect 
to the planetary scattering plane, which is the plane through 
the centers of the star, planet, and the observer.
They represent the following flux differences:
\begin{eqnarray}
    Q & = & F_{0^\circ} - F_{90^\circ} \\
    U & = & F_{45^\circ} - F_{135^\circ},
    \label{eq_U}
\end{eqnarray}
where $F_{x^\circ}$ is the flux measured through a linear 
polarization filter 
with its optical axis making an angle of $x^\circ$ 
with the reference plane, measured by rotating in the clockwise 
direction when looking toward the planet
\citep{Hovenier2004,Hansen1974}.
Fluxes $Q$ and $U$ can be redefined with 
respect to another reference plane, such as the optical plane 
of an instrument, using a rotation matrix 
\citep[see][for the definition]{Hovenier1983}.
The circularly polarized flux $V$ is positive when the observer 
sees the electric vector of starlight that is reflected by a planet
rotating in the anticlockwise direction, and $V$ is negative when the 
observer sees a rotation in the clockwise direction.

Light of a solar-type star can be assumed to be unpolarized
when integrated over the stellar disk. 
This assumption is based on observations of the Sun itself by 
\citet{1987Kemp} and on observations of solar-type stars (active 
and inactive FGK stars) by \citet{2017Cotton}.
Anomalies in the symmetry of the stellar disk, such as flares 
and/or spots, are expected to produce degrees of linear 
polarization of about $10^{-6}$ 
\citep{2015Kostogryz,2011Berdyugina}. 
We represent the stellar flux vector that is incident 
on the planet by the (column) vector 
${\bf F}_0= F_0 [1,0,0,0] = F_0 {\bf 1}$,
where $\pi F_0$ is the total flux as measured in a plane 
perpendicular to the incident direction.

Starlight that has been reflected by a planet is usually 
polarized because it has been scattered by gases 
and aerosols or cloud particles in the planetary atmosphere
and/or has been reflected by the surface below the atmosphere. 
The degree of polarization of this light is defined as 
\begin{equation}
    P_{\rm tot} = \frac{\sqrt{Q^2 + U^2 + V^2}}{F}.
\label{eq:def-pol-tot}
\end{equation}
The Stokes parameter $V$ of sunlight that is reflected by an
Earth-like planet is expected to be very small
\citep[see][and results in this article]{2018A&A...616A.117R,Kawata1978,Hansen1974},
and $P_{\rm tot}$ is therefore usually virtually equal to the 
degree of linear polarization,
\begin{equation}
    P = \frac{\sqrt{Q^2 + U^2}}{F}.
\label{eq:def-pol-lin}
\end{equation}
Because our Earth-like model planets are 
horizontally inhomogeneous because they have continents 
and clouds and are usually asymmetric with respect to the reference plane,
the Stokes parameter $U$ (see Eq.~\ref{eq_U}) is usually not zero. 
It appears to be very small, as we show in 
Sect.~\ref{Sect_Results}.
When $U \approx 0$, the sign of $Q$ indicates
the direction of $P$: if $Q > 0$ ($Q < 0$), 
the direction is parallel (perpendicular) 
to the planetary scattering plane.

\subsection{Radiative transfer algorithm}
\label{Sect_Radiative_Transfer_Algorithm}

We computed the total and polarized fluxes $F$, $Q$, $U$, and $V$, 
and the degree of polarization of starlight that is reflected 
by a model planet using 
the Python-Fortran tool PyMieDAP.\footnote{PyMieDAP can be 
downloaded under the GNU GPL license at: 
https://gitlab.com/loic.cg.rossi/pymiedap}. 
A detailed description of PyMieDAP is given by \citet{2018PyMieDAP}. 
We summarize the parts that are relevant for the interpretation 
of our numerical results. We computed reflected flux vectors 
for spatially resolved model planets. These results apply,
for example, to Solar System planets. By integrating the spatially
resolved fluxes across the planetary disk, we obtain the 
spatially unresolved reflected flux vectors, which apply 
in particular to (future) observations of exoplanets.

For our spatially resolved computations, the 2D 
planetary disk facing the observer is divided into equal--sized,
square pixels. 
We used 100~pixels along the light equator of the planet 
(except for the images shown in 
Figs.~\ref{fig:RGBdisksNoCld}-\ref{fig:RGBdisks}).
The actual planetary surface area that is covered 
by a pixel increases strongly toward the planetary limb.
For the center of each pixel, we computed the 
illumination and viewing angles on the 3D 
spherical planet
\citep[for detailed definitions, see][]{2018PyMieDAP,deHaan1987}:\\

\noindent
{\em Illumination angle $\theta_0$}:  
the angle between the local vertical 
and the direction toward the star.\\

\noindent
{\em Viewing angle $\theta$}: 
the angle between the local vertical 
and the direction toward the observer.\\

\noindent
{\em Azimuthal angle $\phi - \phi_0$}: 
the difference angle between the plane 
containing the local vertical  and the direction of the 
incident light, and the plane containing the local vertical 
and the direction toward the observer. \\

\noindent
Angles $\theta_0$ and $\phi_0$ depend on the planetary phase angle 
$\alpha$, that is, the angle between the center of the star and the 
observer as measured from the center of the planet. For a planetary
orbit with an orbital inclination angle $i$, 
$90^\circ - i \leq \alpha \leq 90^\circ + i$. 
In this paper, we present results for an edge-on orbit, that is,
for $i=90^\circ$, and $0^\circ \leq \alpha \leq 180^\circ$.
For other inclination angles, the appropriate phase angle
range can be chosen from our results.

Given the illumination and viewing geometries, and the 
atmosphere-surface properties for each pixel, we computed
${\bf F}$ (Eq.~\ref{eq:def-stokes-vector}) of the 
reflected starlight for a given pixel $i$, using \citep{Hansen1974}
\begin{equation}
   {\bf F}_i(\lambda,\theta,\theta_0,\phi-\phi_0)= 
           \cos \theta_0 \hspace{0.1cm}
           {\bf R}_{i1}(\lambda,\theta,\theta_0,\phi-\phi_0) 
           \hspace{0.1cm} {\bf F}_0(\lambda),
\label{eq:reflected_vector}
\end{equation}
where ${\bf R}_{i1}$ is the first column of the local reflection
matrix on the planet (only the first column is relevant, because the
incoming starlight is assumed to be unpolarized).
We computed ${\bf R}_{i1}$ using an adding-doubling 
radiative transfer algorithm in Fortran \citep[]{deHaan1987}, 
that fully includes polarization for all orders of scattering.
We computed the reflection by horizontally
inhomogeneous planets by choosing different atmosphere-surface
models for different pixels (see Sect.~\ref{Sect_Exoplanetmodels}).
The reflection of each pixel was computed independently from that
of its neighbours. Given the physical surface areas covered by 
the pixels, 
the contribution of light that scatters from one pixel to another
and then toward the observer is negligible.

Rather than embarking on a separate radiative transfer computation 
for every pixel, we first computed and stored 
the coefficients ${\bf R}_{i1}^m(\theta,\theta_0)$
($0 \leq m < M$, where $M$ is the total number of coefficients) of the 
expansion of ${\bf R}_{i1}(\theta,\theta_0,\phi-\phi_0)$ into a 
Fourier series \citep[see][for details on this expansion]{deHaan1987} 
for the various atmosphere--surface combinations on the model planet. 
We computed and stored the Fourier coefficients at values of 
$\cos \theta_0$ and $\cos \theta$ that coincide with Gaussian 
abscissae, and additionally at $\cos \theta_0= 1$ and $\cos \theta=1$.
Given local angles $\theta_0$, $\theta$, and $\phi-\phi_0$, 
we can efficiently compute the local ${\bf R}_{i1}$ by summing
\citep[see][]{deHaan1987} the Fourier coefficients obtained
by interpolating between the relevant stored coefficients.

The Stokes parameters $Q_i$ and $U_i$ of a locally reflected
vector ${\bf F}_i$ are defined with respect to the local meridian 
plane. To compute the disk-integrated flux vector, we redefined 
them to the planetary scattering
plane before we summed the vectors for all pixels
using the rotation matrix ${\bf L}_i(\beta_i)$, where  
$\beta_i$ is the rotation angle between the two reference planes
for a given pixel $i$, see \citet{Hovenier1983}
for the precise definitions of ${\bf L}_i$ and $\beta_i$.
We computed the disk-integrated flux vector $\pi {\bf F}$, 
as seen from an observer at distance $d$, using
\begin{equation}
   \pi {\bf F}(\lambda,\alpha)= \frac{F_0(\lambda)}{d^2} \sum^N_{i=1} 
        \hspace{0.1cm} \mu_i \mu_{0i}
            {\bf L}_i(\beta_i) \hspace{0.1cm}
            {\bf R}_i(\lambda,\mu_i,\mu_{0i},\phi_i-\phi_{0i}) 
            \hspace{0.1cm} dO_i,
\label{eq:reflected_fluxes}
\end{equation}
where $N$ is the number of pixels, $dO_i$ is the 3D surface area covered
by pixel $i$, and $\mu_i=\cos{\theta_i}$ and 
$\mu_{0i}=\cos{\theta_{0i}}$
are the cosines of the local angles $\theta_i$ and $\theta_{0i}$,
respectively. 
The surface area $dO_i$ is related to the local viewing angle $\theta_i$, 
where $dO_i \mu_i$ equals the pixel size, which we chose to be the same
for all pixels.

We show results at five wavelengths:
350, 443, 550, 670, and 865~nm.
To make our results independent of distances, radii, etc.,
we normalized the reflected flux vectors such 
that the disk-integrated reflected total flux 
presented in this paper at $\alpha=0^\circ$ equals 
the planetary geometric albedo 
\citep[see, e.g.,][]{2018PyMieDAP,stam2008spectropolarimetric}, 
which we denote by '$F$' in what follows.  When
Eq.~\ref{eq:reflected_fluxes} is used, our results can straightforwardly
be scaled to a given planetary system and stellar type.
Depending on the stellar spectrum, 
the planetary color in total and polarized flux would change.
Because the degree of polarization $P$ is a relative measure
(see Eq.~\ref{eq:def-pol-lin}), it would not require normalization
or scaling.

Furthermore, our results pertain
to planets that are spatially resolved from their parent star and
background light of the parent star is not included. To translate 
our results for planets that are spatially unresolved from their star
\citep[for examples for close-in giant planets, see][]{Seageretal00},
the flux vector of the star needs to be added to 
the planetary flux vector (the starlight is usually
unpolarized \citep[][]{1987Kemp}). 
Spatially resolving a planet in the habitable
zone of a solar-type star from its star
requires extreme adaptive optics systems 
and advanced techniques to block the direct starlight. 
Even with these techniques, background starlight remains 
at the position of the planet on the detector. 
However, as it is a differential method, adding polarimetry as a 
technique enhances the achieved contrast by three to five orders of
magnitude \citep[][]{Keller2010}.

We also ignored contributions of starlight scattered by (exo)zodiacal 
dust, interstellar dust, and instrumental effects to avoid 
making our results too specific and 
expanding the parameter space too much. These contributions
might be added, similarly to the background starlight.
Our model planet has a circular orbit around its star,
with an inclination angle of 90$^\circ$.
The obliquity of the planetary rotation axis 
equals that of Earth, $23.45^\circ$, 
and the rotation period is equal to one sidereal Earth day.
The orientation of the planetary rotation axis with respect 
to the parent star and the observer depends on the day of the year at 
full phase. The results of our simulations can be adapted for a 
differently inclined orbit
by choosing the appropriate phase angle range, except that the
distribution of continents and clouds on the model planet 
remains the same (e.g.\ our results do not include a full view
of a polar cap of the model planet).
By adapting the temporal scale of the results, the rotation
period can be adapted, except for the effect that a 
different
rotation period would have on the atmospheric parameters, in
particular, on those of the clouds.

\subsection{Computational costs}

The computational costs of our numerical simulations depend on a 
large number of numerical parameter settings 
\citep[][]{2018PyMieDAP,deHaan1987}.   
For the results presented in this paper, and thus for the data files
made publicly available, we chose the most accurate settings
without considering the computational costs.  
In particular, with 100 pixels across the
equator of the model planet, about 7850 pixels covered the planetary
disk when we modeled the local variations in atmosphere and surface 
properties (see Sect.~\ref{Sect_Atmospheres}). For every 
local atmosphere-surface model, we computed the reflected
Stokes vector (as coefficients of its expansion in a Fourier series,
see \citet{2018PyMieDAP}) for up to 151 different local solar zenith 
angles $\theta_0$ and up to 151 different local viewing
zenith angles $\theta$. These Fourier-coefficient files were stored
in a database for use in the computation of the disk-integrated
Stokes vectors. 
For the latter, we used a temporal resolution for the rotation
of the model planet of 15~minutes, covering 184 Earth days, thus
17,664 disk-integrations (with increasing $\alpha$, the number
of illuminated pixels on the planetary disk decreases, and 
the computational costs decrease in turn). 
With these settings, the computation of a
disk-integrated Stokes vector across all phase angles takes about
18~hours on a 2.7~GHz Intel Core~i5 processor for a single
wavelength.

\section{Atmosphere - surface models}
\label{Sect_Exoplanetmodels}

Here, we describe the local atmosphere and surface properties of the
pixels on our model planet. We assign these 
properties using Earth-observation satellite data. 

\subsection{Model atmospheres}
\label{Sect_Atmospheres}

Our model planets have locally plane-parallel layered atmospheres 
that consist of gaseous molecules and (optionally) cloud particles. 
We assumed an Earth-like gas mixture in hydrostatic equilibrium
(using an Earth-like gravitational acceleration $g$ of 9.81 m s$^{-2}$), 
with a mean molecular mass $m_{\rm g}$
of 29~g/mol. We only considered continuum wavelengths and thus 
ignored gaseous absorption.
Our shortest wavelength of 350~nm is just outside the 
ultraviolet absorption band (the Huggins band) of ozone. 
For an Earth-like amount and vertical distribution of ozone, the 
broadband absorption between about 400 and 650~nm 
(the Chappuis band) decreases the total reflected flux at 
550~nm wavelength, but it does not significantly 
affect the degree of polarization at this wavelength, 
because although in principle,
absorption decreases the amount of multiple scattering of light,
and thus usually increases the degree of polarization of the
reflected light, most of the absorption
by ozone takes place at high altitudes in the atmosphere, where 
the gas density and thus the amount of multiple scattering is 
small.
The strength of absorption bands in reflected light signals
depends on the (wavelength-dependent)
absorption cross-section of the gas and also strongly depends on the
amount and vertical distribution of the ozone molecules, 
and cloud particles. 
To avoid introducing too many Earth-specific parameters, we
decided to ignore it \citep[for spectra that include absorption
by O$_3$ in both the Huggins and the Chappuis bands, see][]{stam2008spectropolarimetric}.

The gas molecules are anisotropic Rayleigh scatterers with a
wavelength-independent molecular depolarization factor $\delta$ of 0.03 
\citep{Hansen1974} and a wavelength-dependent refractive index 
\citep{Ciddor1996}.
The surface pressure $p_{\rm s}$ is 1.0~bar. 
Sample values of the atmospheric gaseous (scattering) optical 
thickness $b_{\rm g}$ are 0.62 at $\lambda=350$~nm and 0.043 at 670~nm.
The atmospheric parameters and their values are given in 
Table~\ref{tab:Atmosoverview}.

We divided the atmosphere into three homogeneous layers.
The gaseous scattering optical thickness of each
layer was computed assuming an exponentially decaying pressure 
within each layer (for the radiative transfer algorithm,
only the optical thickness of the layer is relevant, not the vertical 
distribution of the gas within the layer). If they were present,
the cloud particles were in the middle layer. The cloud layer 
always had a vertical extent of $100$~mb, and the pressure levels 
of the three layers were adapted to accommodate different cloud top
altitudes. Topographical variations were not considered, that is,
the bottom of the atmosphere was at 1 bar everywhere,
and the top was bounded by space.
To limit the computational time and to avoid overcomplicating
the models, 
our model atmospheres do not contain particles other than cloud 
particles, in other words, we ignored small, suspended,
so-called aerosol particles 
such as dust and/or sea-salt particles, and/or smoke due to 
biomass burning.
Most of these aerosol particles show significant horizontal 
variability while having relatively small optical thicknesses 
compared to clouds
\citep[for dust, see, e.g.,][]{Ginoux2001}. The influence of 
various types and optical thicknesses of aerosol could be
part of a subsequent study.

\begin{table}[b]
\small
\caption{Parameters of the model atmosphere.}
\label{tab:Atmosoverview}
\begin{tabular}{l|l|l}
\hline
Parameter and unit & Symbol & Value(s)\\
\hline
Surface pressure [bar] & $p_{\rm s}$ & 1.0 \\
Mean molecular mass [g mol$^{-1}$] & $m_{\rm g}$ & 29.0 \\
Molecular depolarization factor [-] & $\delta$ & 0.03 \\
Acceleration of gravity [m s$^{-2}$] & $g$ & 9.81 \\
Gas optical thickness [-] & $b_{\rm g}(\lambda)$ & 0.62 (350~nm); \\
& & 0.23 (443~nm); \\
& & 0.10 (550~nm); \\ 
& & 0.043 (670~nm); \\ 
& & 0.015 (865~nm) \\ 
Cloud particle size distribution &   & Two parameter gamma \\
\hspace*{3mm} Effective radius [$\mu m$] & $r_{\rm eff}$ & 10; 12.5; 15; 17.5\\
\hspace*{3mm} Effective variance & $v_{\rm eff}$ & 0.1 \\
Cloud particle refractive index & & \\
\hspace*{3mm} Real part & $n_{\rm r}$ & 1.33 \\ 
\hspace*{3mm} Imaginary part & $n_{\rm i}$ & $10^{-8}i$ \\
Cloud optical thickness [-] & $b_{\rm c}(\lambda)$ & 0; 5; 10; 20 \\
Cloud top pressure [mbar] & $p_{\rm c}$ & 500; 700; 850 \\
Cloud vertical extent [mbar] & - & 100 \\
\hline
\end{tabular}
\centering
\end{table}

\begin{table*}[ht]
\tiny
\caption{References of the parameter databases (upper part of the table) and of the surface models (lower part of the table) used for the Earth-like model planet.
}
\label{tab:databases}
\begin{tabular}{l|l|l}
\hline
Parameter & Reference & Comments \\
\hline
Cloud optical thickness & \citet{MODISAQUA+Terraatmoscite} & Satellites: Aqua and Terra. Instrument: MODIS.\\
& & Logarithmic daily mean data set (see \citet{hubanks2015modis}; \\
& & \citet{Oreopoulos2007}). \\
& & Data range: January 1, 2011 - December 31, 2011. \\
& & Temporal resolution: one day.\\
& & Spatial resolution: 1.0$^\circ$x1.0$^\circ$.\\
Cloud particle effective radii & \citet{MODISAQUA+Terraatmoscite} & Satellites: Aqua and Terra. Instrument: MODIS. Daily mean data set. \\
& & Data range: January 1, 2011 - December 31, 2011. \\
& & Temporal resolution: one day.\\
& & Spatial resolution: 1.0$^\circ$x1.0$^\circ$.\\
Cloud top pressure & \citet{MODISAQUA+Terraatmoscite} & Satellites: Aqua and Terra. Instrument: MODIS. Daily mean data set. \\
& & Data range: January 1, 2011 - December 31, 2011. \\
& & Temporal resolution: one day.\\
& & Spatial resolution: 1.0$^\circ$x1.0$^\circ$.\\
Surface cover type & \citet{MODISLandcite} & Satellites: Combined Aqua and Terra. Instrument: MODIS.\\
& & Data range: January 1, 2011 - December 31, 2011. \\
& & Temporal resolution: Yearly.\\
& & Spatial resolution: 0.05$^\circ$x0.05$^\circ$.\\
Snow cover & \citet{MODISAQUASNOW} & Satellite: Aqua. Instrument: MODIS.\\
& & Data range: January 1, 2011 - December 31, 2011. \\
& & Temporal resolution: Daily.\\
& & Spatial resolution: 0.05$^\circ$x0.05$^\circ$.\\

Wind speed above the ocean & \citet{CCMP2011data,CCMP2011} & The input data are a combination of intercalibrated satellite data from \\
& & numerous radiometers and scatterometers and in situ data from moored \\
& & buoys. \\
& & Data range: January 1, 2011 - December 31, 2011. \\
& & Wind speed altitude: 10 meters. Temporal resolution: 6 hours.\\
& & Spatial resolution: 0.25$^\circ$x0.25$^\circ$, near globally.\\
Pale soil albedo & \citet{Tilstra2017} &   Satellite: ENVISAT. Instrument: SCIAMACHY.\\
& & Lambertian-equivalent reflectivity data set.\\
& & Temporal resolution and coverage: one month in 2002-2012.\\
& & Spatial resolution: 1.0$^\circ$x1.0$^\circ$.\\
& & Soil albedo based on the mean spectral reflectivity over a full year\\
& & and over a slab of desert sands in Egypt, with the coordinates\\
& & 26.25$^\circ$-27.25$^\circ$ latitude, 26.25$^\circ$-27.25$^\circ$ longitude.\\
Dark brown soil albedo & \citet{Ecostressprep,Aster2009} & Retrieved from https://speclib.jpl.nasa.gov\\
& & Library developed as part of the ASTER and ECOSTRESS projects.\\
Ice and snow albedo & \citet{Ecostressprep,Aster2009} &  Same comment as provided for dark brown soil albedo.\\
Deciduous forest albedo & \citet{Ecostressprep,Aster2009} & Same comment as provided for dark brown soil albedo.\\
Grass albedo & \citet{Ecostressprep,Aster2009} &  Same comment as provided for dark brown soil albedo.\\
Steppe albedo & \citet{Steppealbedo} &  Spectral albedo retrieved from a patch of ground in the Southern\\
& & Great Planes, USA, using a multifilter radiometer at 25 m altitude. \\
& & The patch of ground consisted of ~56$\%$ vegetation, the remaining \\
& & part was soil. The measurement was taken on June 20, 2011.\\
Scattering coefficients of soil & \citet{moreno2006scattering} & Retrieved from the Amsterdam light-scattering \\
& & database \citep{dustdatabase2012}. \\
\hline
Surface model & Reference & Comments \\
\hline
Ocean & \citet{TreesandStam2019} & Rough Fresnel-reflecting interface, subinterface pure seawater body, \\
& & ocean bottom, and wind-generated foam.\\
Deciduous forest & \citet{Maignan2009,Roujean1992} & Combined vegetation model, which includes the complete reflection \\
& & matrix for the total reflection and linearly polarized reflection. \\
& & Circularly polarized reflection not included. \\
Grass & \citet{Maignan2009,Roujean1992} & Combined vegetation model, which includes the complete reflection \\
& & matrix for the total reflection and linearly polarized reflection. \\
& & Circularly polarized reflection not included. \\
Steppe & \citet{Maignan2009,Roujean1992} &  Combined vegetation model, which includes the complete reflection \\
& & matrix for the total reflection and linearly polarized reflection. \\
& & Circularly polarized reflection not included. \\
Desert & Amsterdam light-scattering database & Pale soil. Optically thick dust layer ($b = 100 $) just above a black surface.  \\
& \citep{dustdatabase2012} & Mie scattering matrix provided at 663~nm. See text for details.\\
Shrublands & Amsterdam light-scattering database & dark brown soil. Reflection model the same as that of the desert, \\
& \citep{dustdatabase2012} & but scaled for the albedo of dark brown soil. See text for details.  \\
Ice or snow & - & Lambertian-reflecting surface with appropriate ice or snow albedo.  \\
\hline
\end{tabular}
\centering
\end{table*}

The spatial and temporal variability in the cloud cover
is derived from observations by MODIS, the Moderate Resolution 
Imaging Spectroradiometer \citep[for a detailed description of the 
MODIS data sets, including retrieval algorithms, 
see][]{EOSMODISvolI,EOSMODISvolII}.  
We describe a cloud by the pressure of its top, $p_{\rm c}$,
its optical thickness, $b_{\rm c}$, and the cloud particle
size distribution and composition. 

Each cloudy model pixel has a cloud fraction. This is zero for 
a cloud-free pixel. Even with as many as 100 pixels along the  
light equator of the model planet, the pixels are much larger than 
typical horizontal variations in clouds, and they are much larger 
than MODIS ground pixels. 
For a mixture of cloud-free and partly and fully cloudy
MODIS ground
pixels within a model pixel, we computed the locally reflected 
Stokes vector for a cloud-free and a cloudy model pixel,
and took their weighted sum to compute the reflected 
Stokes vector for the pixel as a whole. 
In this way, we accounted for 
the cloud fraction as reported for all cloudy MODIS ground
pixels. The mean cloud fraction of our model planet is about 66~\%.

The MODIS database provides a cloud top pressure for all of its
pixels, and we computed $p_{\rm c}$ of a cloudy model pixel by taking 
the average of the cloud top pressures of the MODIS pixels 
within our model pixel, weighted by their cloud fractions.
To limit the number of different atmosphere-surface models, we 
divided $p_{\rm c}$ for the model pixels into three categories:
500~mbar for $0 \leq p_{\rm c} < 600$~mbar,
700~mbar for $600 \leq p_{\rm c} < 800$~mbar, and
850~mbar for $p_{\rm c} \geq 800$~mbar
(see Table~\ref{tab:Atmosoverview}).

We computed $b_{\rm c}$ for a cloudy model pixel by taking the
mean of the cloud optical thickness values of the MODIS ground pixels 
within the model pixel. 
In particular, we take the mean of the MODIS 
logarithmic daily mean data set (see Table~\ref{tab:databases}). 
This data set provides the cloud optical thickness for a given 
MODIS pixel $i$ 
as $\log{b_{{\rm c}i}}$, where $b_{{\rm c}i}$ is the cloud 
optical thickness. 
When we computed the mean $b_{\rm c}$, we weighted $\log{b_{{\rm c}i}}$ 
of each MODIS pixel with its cloudiness fraction. 
This logarithmic mean as a good approximation for the 
cloud optical thickness has been proposed by 
\citet{hubanks2015modis} and \citet{Oreopoulos2007}.

The wavelength at which a cloud optical thickness is provided in 
the MODIS data set depends on the surface: 
$b_{\rm c}$ pertains to 645~nm above a dry surface, 
to 858~nm above an ocean,
and to 1240~nm above snow or ice. 
We obtain $b_{\rm c}$ at another wavelength by multiplying the 
value from the data set with the ratio of 
the cloud particle extinction cross-section at that other wavelength 
and at the given wavelength.
The cloud optical thickness depends somewhat on the
wavelength: for a cloud with $b_{\rm c}=10$, 
composed of particles with $r_{\rm eff}=12.5 \mu$m and $v_{\rm eff}=0.1$, 
the cloud optical thickness, for example, ranges from 9.775 to 10.002,
from 350 to 865 nm.
The extinction cross-sections of the cloud particles were
computed using a Mie-algorithm based on \citet{de1984expansion}.
To limit the number of different atmosphere-surface models,
we divided $b_{\rm c}$ for the model pixels into four categories:
0.0 when $b_{\rm c} < 0.01$, 5.0 when $0.01 \leq b_{\rm c} < 7.5$, 
10.0 when $7.5 \leq b_{\rm c} < 15.0$,
and 20 when $b_{\rm c} >$ 15,
at 645, 858, or 1240~nm, depending on the surface 
(see above).

The cloud particles were assumed to be homogeneous, liquid-water spheres, 
with a refractive index of $n= 1.33 + 10^{-8}i$ \citep{Hale:73}. 
Their sizes are described by a two-parameter gamma distribution
\citep{Hansen1974}, with an effective particle
radius $r_{\rm eff}$ and an effective variance $v_{\rm eff}$.
We used $v_{\rm eff}=0.1$
\citep{ha06800b,nakajima1990determination}.
The cloud particle effective radius was provided by the MODIS
database, and we computed $r_{\rm eff}$
for a model pixel by taking the average of the effective radii
of the MODIS pixels within our model pixel, weighted with the
cloud fractions.
We computed the single-scattering properties (i.e.\ single-scattering
matrix, albedo, and extinction cross-section) of the cloud particles
using a Mie algorithm based on \citet{de1984expansion}.

To limit the number of different atmosphere-surface models, 
we divided $r_{\rm eff}$ for the model pixels into four categories 
\citep[see][]{nakajima1990determination}: 
10~$\upmu$m for 0.0~$\leq r_{\rm eff} \leq 11.25~\upmu$m, 
12.4~$\upmu$m for 11.25~$< r_{\rm eff} \leq 13.75~\upmu$m
15.0~$\upmu$m for 13.75~$< r_{\rm eff} \leq 16.25~\upmu$m, and
17.5~$\upmu$m for $r_{\rm eff} > 16.25~\upmu$m.

\subsection{Model surfaces}
\label{Sect_Surfaces}

The surface of our model Earth was covered by four surface types: 
ocean, vegetation, soil, and snow or ice. Vegetation was subdivided into
deciduous forest, grass, and steppe. Soil was subdivided into
sand desert and shrubland.
For  the  spatial  and  temporal  variability  in  surface type 
we again used MODIS data.
When the MODIS data showed various surface types within a model
pixel, we assigned the most abundant surface type to the pixel. 
The different surface types and their reflection properties are 
described in more detail below. \\

\noindent{\em Ocean} \\
\noindent 
The ocean consisted of a Fresnel-reflecting and 
transmitting air-water interface with water below. 
The interface was rough due to wind-driven waves that were 
described by randomly oriented flat facets. The standard 
deviation of the wave-facet inclinations increased with the 
wind speed, widening the sun-glint pattern on the surface 
\citep{mishchenko1997satellite,cox1954measurement}. 
We included scattering of light within the water body 
by describing it as a stack of horizontally homogeneous layers of 
pure seawater and by computing its reflection using an 
adding-doubling algorithm \citep[]{deHaan1987}. 
The sub-ocean surface was black.

The single scattering in pure seawater is described by anisotropic
Rayleigh scattering \citep{Hansen1974} with a depolarization factor 
of 0.09 \citep{Chowdharyetal2006,morel1974optical}, combined with
realistic absorption and scattering coefficients for pure seawater
\citep{SogandaresandFry1997,pope1997absorption,SmithandBaker1981}.
This results in a natural blue (for low chlorophyll 
concentrations) water body. 
The real part of the refractive index of the water was assumed to be 1.33
\citep{Hale:73} and wavelength independent, 
which does not significantly influence the ocean reflection.
Furthermore, we took the reflection by wind-generated surface foam 
into account 
\citep[see][]{koepke1984effective} and corrected for the 
energy surplus across the rough air-water interface at grazing 
angles that was caused by neglecting wave shadowing 
\citep[see also][]{zhaietal2010,tsang1985theory}.

To limit the number of different atmosphere-surface models, 
we divided the ocean surface wind speed $v$ for the model pixels
into two categories: 5~m/s for $v < 6$~m/s, 
and 7~m/s for $v \geq 6$~m/s.
This division is based on the global variation of mean wind speed 
measurements between January 1, 2011, and December 31, 2011, 
at an altitude of 10 meters 
\citep[see][]{CCMP2011}\footnote{Cross-calibrated multi-platform
(CCMP) wind vector analysis, 
http://www.remss.com/measurements/ccmp/}.
The parameters of the ocean properties are
listed in Table~\ref{tab:oceanmodel}.
For a detailed description of our ocean reflection algorithm, 
see \citet{TreesandStam2019}. \\

\begin{table}[t]
\caption{Parameters of the ocean.}
\label{tab:oceanmodel}
\begin{small}
\begin{tabular}{l|l|l}
\hline
Parameter and unit &     Symbol       &     Value  \\ \hline
Wind speed {[}m/s{]} & $v$          & 5.0; 7.0     \\
Foam albedo & $a_{\rm foam}$          & 0.22     \\
Depolarization factor water  & $\delta_{\rm water}$ & 0.09  \\
Refractive index above the air-water interface   & $n_1$       & 1.0   \\
Refractive index below the air-water interface & $n_2$       & 1.33  \\
Chlorophyll concentration {[}mg/m$^3${]}   & {[}Chl{]}  & 0.0    \\
Ocean depth {[}m{]} & $\Delta z_{\rm ocean}$          & 100.0   \\ 
Ocean bottom surface albedo   & $a_{\rm ocean bottom}$  & 0.0 \\
\hline
\end{tabular}
\end{small}
\centering
\end{table}

\noindent{\em Vegetation} \\
\noindent
A vegetated surface on our model planet 
can be covered by deciduous forest, grass, or steppe.
We used a combination of surface reflection models to describe
the bidirectionally reflected total and linearly polarized fluxes.
We also accounted for the reflection of incident linearly polarized light,
as the diffuse skylight is usually linearly polarized. 
We ignored the reflection of incident circularly 
polarized light, however, as this signal is very weak
\citep[][]{2018A&A...616A.117R}.
We also ignored the circular polarization of light that is reflected
by vegetation. This light is expected to be slightly 
circularly polarized \citep{2019Patty,2009Sparks},
but to our knowledge, there is no numerical 
bidirectional reflection model for vegetation that includes 
circular polarization (the measurements by \cite{2019Patty} 
cover a range of incidence and reflection angles that is, as yet too 
limited to convert the data into a fully bidirectional reflection model).

To describe the bidirectionally reflected total flux, we used the 
numerical model developed by \cite{Roujean1992},
\begin{eqnarray}
   \rho(\theta,\theta_0,\phi-\phi_0,\lambda) = & 
       \hspace*{-0.8cm} k_0(\lambda) + k_1(\lambda) \hspace*{0.1cm} 
       f_1(\theta,\theta_0,\phi-\phi_0) + 
       \nonumber \\
       & \hspace*{1.5cm} + \hspace*{0.1cm} 
       k_2(\lambda) \hspace*{0.1cm} 
           f_2(\theta,\theta_0,\phi-\phi_0),
\end{eqnarray}
where $\rho$ is the bidirectional reflectance. 
The function $f_1$ is the so-called geometric component of the reflection,
which includes the influence of shadows (mutual shadows,
such as one tree casting a shadow on another tree, are ignored),
and $f_2$ is the volume component, which includes the amount of 
local reflection (e.g.\ the density of leaves in a tree canopy)
\citep[for details, see][]{Roujean1992}. 
Functions $f_1$ and $f_2$ vanish at $\theta = \theta_0 = 0^\circ$.
Using remote-sensing data, 
\cite{Roujean1992} derived parameters $k_0$, $k_1$, and $k_2$ 
for various types of vegetation and two wavelength bands:
580~to 680~nm and 730~to 1100~nm.

To account for the spectral variability of the reflection within 
these bands, in particular the green bump and red-edge features, 
the $k$-parameters were scaled to the appropriate hemispherical 
albedo of the vegetation type considered.
The hemispherical albedo of deciduous forest and grasslands was
taken from the ECOSTRESS spectral library 
\citep[][]{Ecostressprep,Aster2009}, 
and that of steppe
from the Atmospheric Radiation Measurement research facility 
\citep[][]{Steppealbedo}
(see Table~\ref{tab:databases}). 

To account for the reflected linearly polarized fluxes,
we used a modified version of the numerical model developed by 
\cite{Maignan2009}, following \citet{Schaepman-Strub_2006}, 
to obtain a bidirectional polarization distribution function (BPDF).
This BPDF is a linear one-parameter-model, a simplification of 
the nonlinear, two-parameter model by \cite{Nadal_1999}.
To account for different types of vegetation, 
the BPDF uses the normalized difference vegetation index (NDVI) 
and a so-called $\alpha$-parameter. 
A table with these parameters for several types of vegetation 
is provided in \cite{Maignan2009}.
With the NDVI parameter, the model of \citet{Maignan2009} accounts 
for the spectral variation in total flux,
and with the $\alpha$-parameter, 
a best fit with observations of the vegetation type considered is achieved 
\citep[see][for a detailed description]{Maignan2009}.
\\

\noindent{\em Arid regions} \\
\noindent
We used two types of arid regions: sandy deserts, and shrublands. 
For these surface types, we assumed the same reflection functions,
but different spectral albedos. 
We approximated the slightly polarized bidirectional reflection 
using the reflection matrix of an optically thick 
(i.e.\ $b_{\rm ext}=100$) layer of irregularly shaped mineral particles. 
This matrix was computed using our adding-doubling algorithm. 
For the single-scattering matrix of the mineral particles, we used 
that of the olivine S particles 
\citep[][]{dustdatabase2012,moreno2006scattering}, while we 
scaled the single-scattering albedo of the mineral particles
to fit the hemispherical albedo of 
pale soil for the sandy desert, and dark brown entisol for the 
shrubland. The spectral albedos and databases we used are provided in 
Table~\ref{tab:databases}.\\

\noindent{\em Ice or snow cover} \\
\noindent
We described fields of snow 
and ice on land and inland waters as Lambertian reflectors. Their 
reflection is thus isotropic and unpolarized.
A flat, smooth ice sheet 
free of rocks, snow or other rough materials might also be 
approximated as a Fresnel reflector. 
However, given the physical dimensions of our model pixels,
and thus their spatial extent on the surface,
a Lambertian approximation appears to be fitting well 
\citep{coakley2003reflectance}.
To account for temporal changes in snow-covered land and inland water, 
we applied the daily global snow cover mask of MODIS on
the yearly surface cover type data set. 
We did not account for sea ice.
The names of the spectral albedo and surface type databases 
are provided in Table~\ref{tab:databases}.

\section{Results}
\label{Sect_Results}

\subsection{RGB-images}
\label{Sect_Results_RGB}

\begin{figure*}[ht!]
\centering
\includegraphics[width=0.94\textwidth]{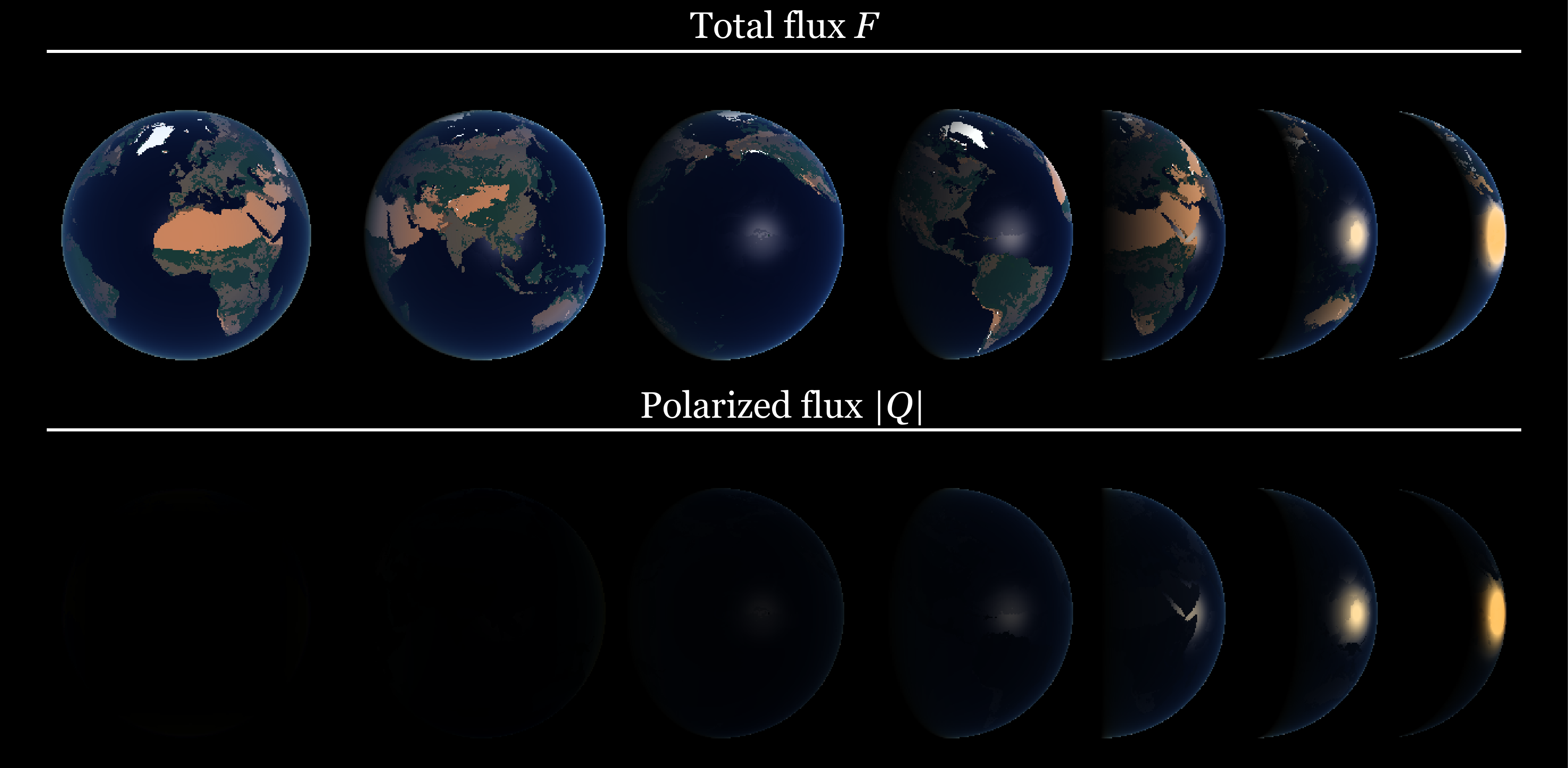}
\caption{Our model planet without clouds in $F$ (top) and $|Q|$ (bottom), 
         using MODIS data of the following days (from left to right): 
         July 1, July 28, August 24, September 20, October 17, 
         November 13, and December 10, 2011, at various sub-observer 
         longitudes and the following phase angles $\alpha$: 0$^\circ$, 
         20.41$^\circ$, 40.82$^\circ$, 60.25$^\circ$, 90.49$^\circ$,
         109.92$^\circ$, and 135.25$^\circ$. The images are 
         not a temporal sequence, they only illustrate various 
         appearances of the planet.
         The RGB colors were computed using weighted additive 
         color mixing of the fluxes at $\lambda=443$~nm (blue), 
         550~nm (green), and 670~nm (red), such that when these 
         fluxes are equal, the resulting color is white.
         The grayscale of each pixel is computed from the sum of 
         the fluxes at the three wavelengths.
         These images have 220 pixels along the light equator
         of the planet.
         }
\label{fig:RGBdisksNoCld}
\end{figure*}

\begin{figure*}[ht!]
\centering
\includegraphics[width=0.94\textwidth]{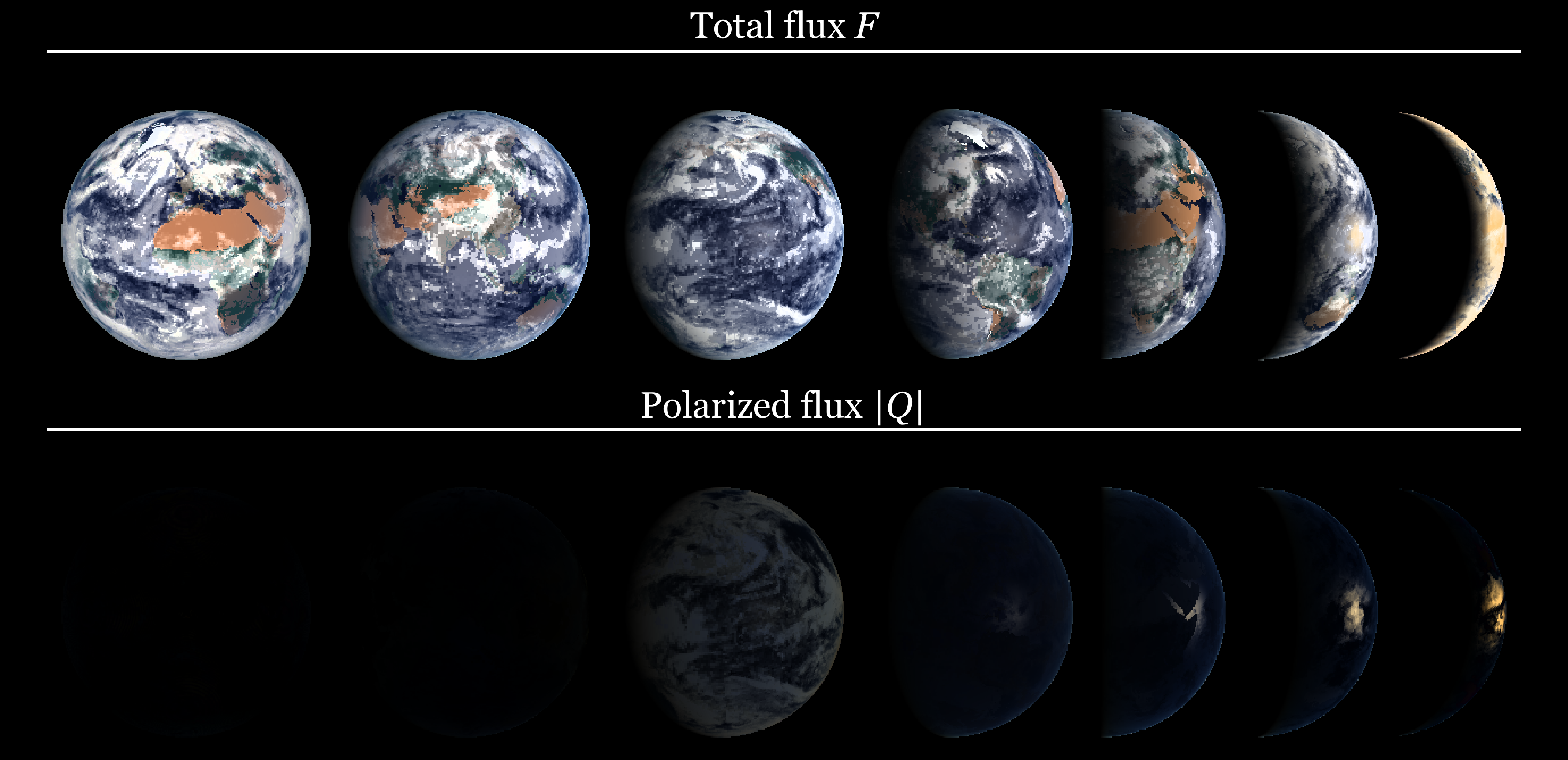}
\caption{Similar to Fig.~\ref{fig:RGBdisksNoCld}, but with cloud
         properties and patterns from the MODIS data of the respective 
         days.}
\label{fig:RGBdisks}
\end{figure*}

Figures~\ref{fig:RGBdisksNoCld} and~\ref{fig:RGBdisks} show 
RGB images of our model Earth without and with clouds
at various orientations and phase angles. 
These figures illustrate the variation in RGB colors of 
the total and polarized fluxes $F$ and $|Q|$ across the model planet.
The model parameters are as described in Sect.~\ref{Sect_Exoplanetmodels}.
The RGB images clearly show that the disk-integrated fluxes are expected
to vary spectrally and temporally as the planet rotates and orbits
its star. A particularly striking feature on the cloud-free disks
is the glint, that is, the sunlight that is reflected by the ocean,
the contribution of which to the signal of the disk 
increases with increasing phase angle $\alpha$. 
The glint can also be seen 
on the cloudy disks, but it is more diffuse, as clouds hide the ocean
and its reflection from view. 
The reddening of the glint in $F$ and $|Q|$ with increasing $\alpha$
has been described by \citet{TreesandStam2019}. These authors also
argued that the reddening in $|Q|$ is indicative of the presence of
a cloud-free or cloudy ocean, while $F$ also reddens when the planet
has no ocean, but only clouds.
The slight increase in $|Q|$ of the cloudy model planet at 
$\alpha=40^\circ$ is due to the primary rainbow, that is, light that 
has been reflected inside the cloud droplets once 
\citep[see, e.g.,][]{karalidi2012looking,stam2008spectropolarimetric,bailey2007rainbows}.


\subsection{Phase curves}
\label{Sect_Results_phase}

\begin{figure*}
\centering
\includegraphics[width=0.91\textwidth]{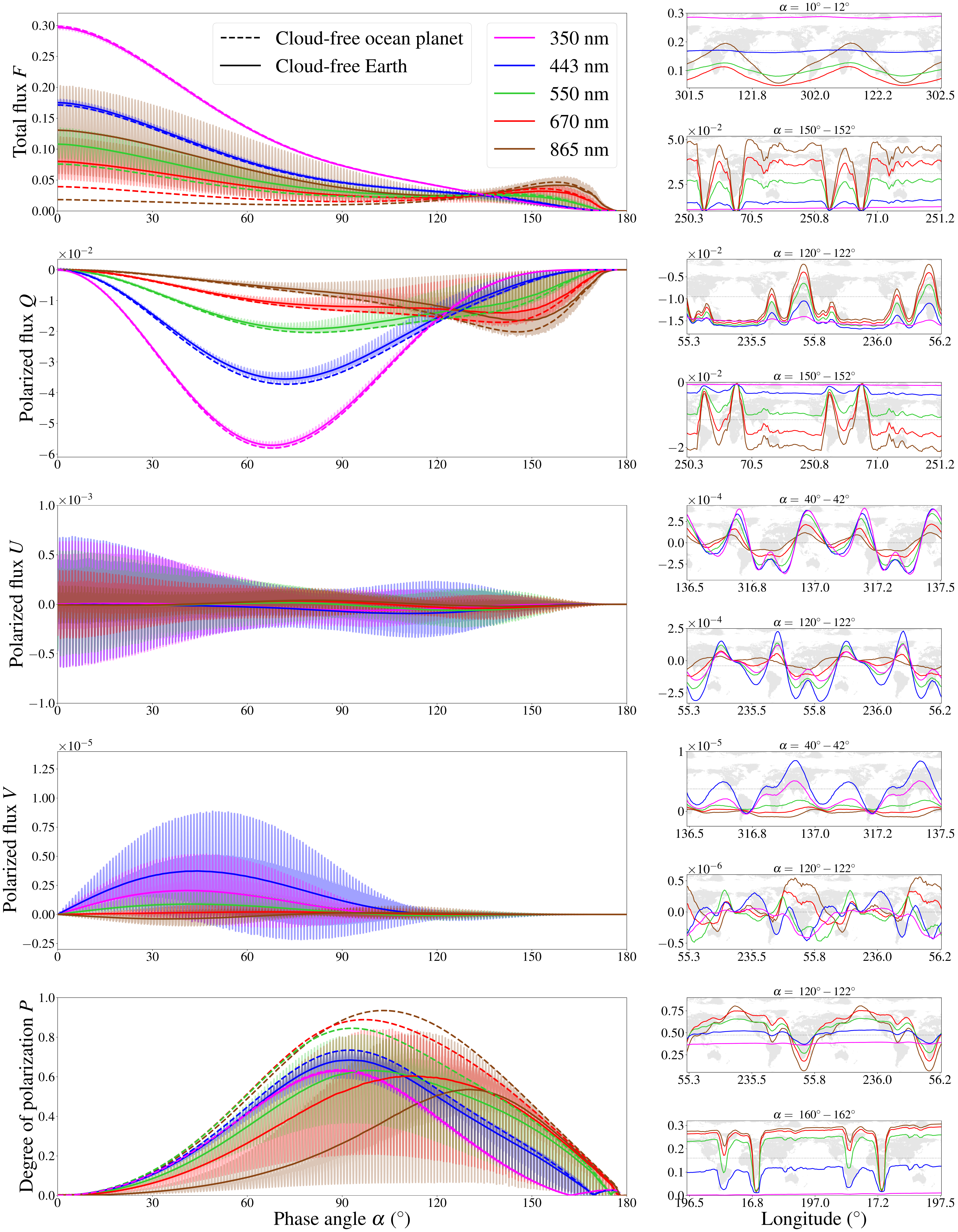}
\caption{Numerical results for the cloud-free model Earth. 
         Left: Total flux $F$, linearly polarized fluxes $Q$ and $U$, 
         circularly polarized flux $V$, and degree of polarization $P$ 
         as functions of $\alpha$ at $\lambda=350$~nm (pink lines),
         443~nm (blue), 550~nm (green), 670~nm (red), and 865~nm (brown). 
         The temporal resolution of the thin lines is 15~minutes.
         The thick solid lines are the diurnal (24h) averages.
         At $\alpha=0^\circ$, the time is July 1, 12:00 GMT, 2011.
         At $\alpha=180^\circ$, the time is December 31, 11:45 GMT, 
         2011. The thick dashed lines are the diurnal averages of a 
         cloud-free 'ocean planet' \citep[cf. Fig.~1 of][]{TreesandStam2019}, 
         thus of a planet without clouds and continents. Right:
         Close-ups of the 15-minute time-resolution phase curves for 
         two phase-angle regions as functions of the longitude in the
         middle of the longitude range that is visible to the observer.
         The background of each close-up shows the corresponding 
         geographic map of Earth. Note the different vertical scales.}
\label{fig:PhaseCurvesNoClouds}
\end{figure*}

\begin{figure*}[!ht]
\centering
\includegraphics[width=0.91\textwidth]{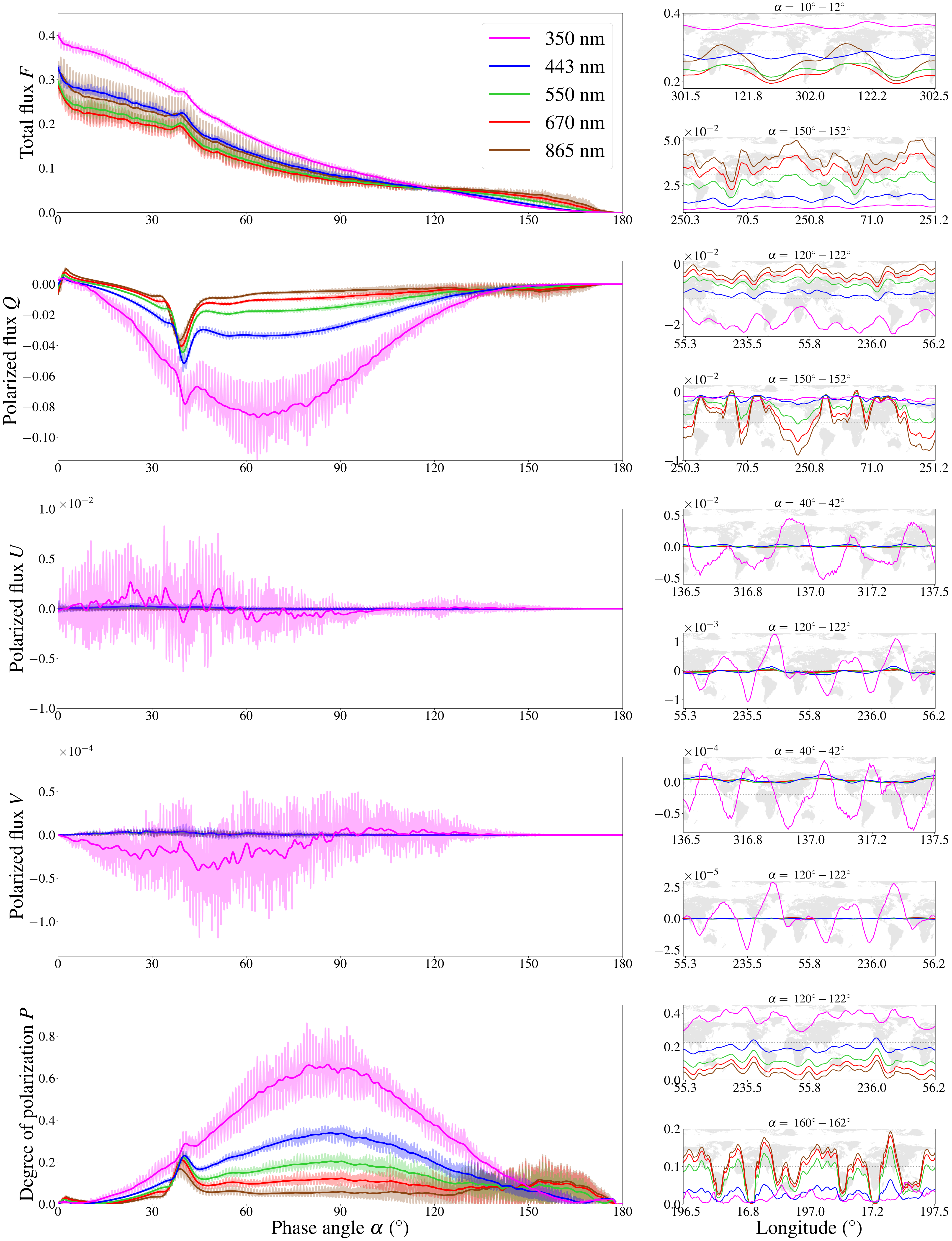}
\caption{Similar to Fig.~\ref{fig:PhaseCurvesNoClouds}
         but with clouds.}
\label{fig:PhaseCurves}
\end{figure*}

Figure~\ref{fig:PhaseCurvesNoClouds} shows $F$, $Q$, $U$, $V$, and the 
degree of polarization $P$ of light reflected by the rotating 
cloud-free model planet as functions of $\alpha$, and for $\lambda$ 
ranging from 350 to 865~nm. 
The simulations start at $\alpha=0^\circ$ on July~1, 12:00~GMT, 
2011 (the geographic longitude centered at the illuminated and visible 
disk is 0$^\circ$, like in 
Figs.~\ref{fig:RGBdisksNoCld}-\ref{fig:RGBdisks}), and stop at 
$\alpha=180^\circ$ on December~31, 11:45~GMT, 2011. 
The figure also includes curves for a cloud-free 
model planet whose surface is completely covered by ocean
(the 'ocean planet' phase curves of Fig.~1 of \citet{TreesandStam2019})
with a surface wind speed $v$ of 7~m/s,
instead of by a mixture of oceans and continents.
Figure~\ref{fig:PhaseCurves} is similar to 
Fig.~\ref{fig:PhaseCurvesNoClouds}, but has clouds.
In Figs.~\ref{fig:PhaseCurvesNoClouds} and~\ref{fig:PhaseCurves},
the thin lines were computed with a temporal resolution of
15~minutes, while the thicker, smoother lines are the averages over
24~hours ($\alpha$ increases by about 1$^\circ$ in this 
period). 


The two figures show that $|U|$ is about a factor 
100 smaller than $|Q|$ across most of the phase angle range
because our reference plane coincides with
the light equator of the planet, and therefore any polarized flux 
$|U|$ is due to
secondary and higher-order scattered light, which has a low
degree of polarization when integrated across the disk, 
even in the presence of horizontal inhomogeneities.
The polarized flux $|U|$ is highest at the shortest wavelengths because
because there, the atmospheric gaseous optical thickness is highest
there and thus more multiple scattering takes place. This is especially
apparent for the cloudy model planet, where multiple scattering
in the gaseous atmosphere above the clouds produces much more
$|U|$ at 350~nm than at 443~nm, where the gas optical thickness
above the clouds is far lower.

The circularly polarized flux $|V|$ is a factor of about 10$^4$ lower
than $Q$. This flux only arises for light that has been scattered at 
least twice, including at least once by the cloud particles,
as Rayleigh scattering alone does not produce circular polarized fluxes
\citep[see][and references therein]{2018A&A...616A.117R}. 
When a model planet is mirror-symmetric with respect to
the planetary scattering plane, the circularly polarized flux of
the hemisphere above the scattering plane (the northern hemisphere) 
will be equal to that of the hemisphere below the scattering plane 
(the southern hemisphere), except for the sign:
because of the mirror symmetry in the scattering geometries,
$V$ of the northern hemisphere equals -$V$ of the southern
hemisphere \citep[see][for comparable phase curves of $V/I$ as 
computed for the two hemispheres of the Earth, except with a
black surface to model the ocean, while our ocean is modelled
as a rough, Fresnel reflecting surface with scattering in the 
water body below]{Munoz2015Towards}. Integrated over the planetary disk,
$V$ of a mirror-symmetric planet will thus equal zero. For a 
horizontally inhomogeneous planet, such as our model planets, the
disk-integrated $|V|$ is usually not zero, but it is still very small, 
and it decreases with $\lambda$ as the atmosphere's Rayleigh scattering 
optical thickness decreases with the wavelength.
The circular polarization signature of
light scattered by chiral molecules, such as those found
e.g.\ in the cells making up Earth-like vegetation, 
on the northern hemisphere will have the same sign
as that on the southern hemisphere, and will thus not
be cancelled when integrated across the planetary disk,
even though the total signal will still be small
because the fraction of circularly polarized flux
that is reflected by chiral molecules is very small
\citep[see][]{2019Patty}.

Figures~\ref{fig:PhaseCurvesNoClouds} and~\ref{fig:PhaseCurves}
show that except at $\alpha > 120^\circ$, the average (24h) total 
and polarized fluxes decrease with increasing $\lambda$ below 670~nm. 
The reason is that the Rayleigh scattering optical 
thickness in the atmosphere decreases with $\lambda$ (our model atmospheres contain
no ozone, otherwise the fluxes at 350~nm would have been much 
lower than shown here), and because the dark ocean suppresses the
terrestrial average surface albedo. The total flux $F$ at 865~nm
is an exception:
it is higher than at 670~nm (for $\alpha<120^\circ$). 
The reason is that at 865~nm, the albedo of 
vegetation is higher than at 670~nm. The increase in this
albedo with increasing $\lambda$ above about 670~nm
is usually referred to as the red edge
\citep{montanes2006vegetation,2005Seager}. 
Because of the low-polarization signature of the vegetation,
$|Q|$ at 865~nm is lower than at 670~nm.

For $\alpha>120^\circ$, the average (24h) cloud-free model planet 
changes color from blue through white to red because of the presence 
of the Fresnel-reflecting ocean \citep[see Fig.~1 in][]{TreesandStam2019}.
The contribution of the ocean glint (which itself is spectrally neutral) 
increases with increasing $\lambda$ as the
Rayleigh scattering optical thickness in the atmosphere decreases. 
\citet{TreesandStam2019} showed that in $F$, this color change 
may also result from Rayleigh-scattering gas above the clouds, 
that is, also in the absence of an ocean.
The diurnal average of $P$ changes color at $\alpha \sim 144^\circ$ 
and the diurnal average of $Q$ changes color at $\alpha \sim 138^\circ$.
\citet{TreesandStam2019} postulated that the color change of $P$ from 
blue through white to red with increasing $\alpha$ identifies a 
partly cloudy exo-ocean, while the color change in $Q$ 
uniquely identifies an exo-ocean, 
independent of the surface pressure or cloud fraction. 

Compared with the results in \citet{TreesandStam2019}, where all
model planets were ocean planets, Figs.~\ref{fig:PhaseCurvesNoClouds} 
and~\ref{fig:PhaseCurves} show that
when the model planet has oceans and continents, $P$ and $Q$ still 
change color, and their color change is thus still indicative of
the presence of an ocean. On Earth, 71\% of the surface is 
covered by ocean, and in the equatorial regions where the glint occurs,
the coverage is even higher. 

The average (24h) $F$, $Q$, and $P$ phase curves of the cloudy planet
(Fig.~\ref{fig:PhaseCurves}) show the familiar features that
are due to the scattering of light by the cloud particles, 
in particular the primary rainbow near $\alpha=40^\circ$ 
that is due to light that has been reflected once within the 
spherical liquid-water droplets
\citep[][]{karalidi2011flux,stam2008spectropolarimetric,bailey2007rainbows},
an enhancement that was also identified in Fig.~\ref{fig:RGBdisks}.
Fluxes $U$ and $V$ also appear to have a feature at the primary 
rainbow angle, although not very obvious. 
At phase angles below 5$^\circ$, the curves 
show a small enhancement that is due to the glory formed by light
that is back-scattered by the spherical cloud droplets
\citep[see, e.g.,][]{Hansen1974,1974SciAm.231a..60B}.
At such small phase angles, an exoplanet would be extremely
close to its star. 

The close-ups of the diurnal variations (the 15 min curves) in
Fig.~\ref{fig:PhaseCurvesNoClouds} show patterns that repeat about 
every 1$^\circ$ in $\alpha$ and are caused by continents and oceans that 
rotate in and out of view. In Fig.~\ref{fig:PhaseCurves} the clouds add 
variation to the modulation through the surface features and thus 
suppress the repeated patterns. The temporal variation in $F$, 
$Q$, and $P$ increases with increasing $\alpha$ because with decreasing
illuminated area on the planetary disk, local variations in surface and
cloud coverage contribute more prominently to the reflected signal, as
also described by \cite{stam2008spectropolarimetric}, for example.
With increasing $\alpha$, the contribution of the ocean glint to $F$ 
and $Q$ also increases (cf.\ Fig.~\ref{fig:RGBdisks}). At short
wavelengths, the peak-to-peak variability in $F$ and $Q$ of the 
cloud-free planet (Fig.~\ref{fig:PhaseCurvesNoClouds}) is small 
because the Rayleigh scattering optical thickness of the atmosphere is 
high and suppresses the surface signals. With increasing $\lambda$, 
the Rayleigh scattering decreases and the peak-to-peak variations due 
to surface albedo variations are increasingly visible.

In the presence of clouds (Fig.~\ref{fig:PhaseCurves}), $Q$ and 
the very low fluxes $U$ and $V$
show the strongest peak-to-peak 
variability at the shortest wavelength (350~nm). 
The temporal variations in $U$ and $V$ appear to
be inversely correlated at small $\alpha$ and correlated
at larger $\alpha$, which can probably be explained by the fact that 
both $U$ and $V$
are due to multiple scattered light. The precise relation
requires further investigations, however. 
This strong 
variability is correlated with the variability of the clouds: without
clouds, the polarized fluxes are due to incident sunlight that is
scattered upward by the atmospheric gases. Clouds strongly increase 
the total flux that is reflected by a pixel, and while $P$
of the flux reflected by the cloud is usually low,
part of this reflected light is scattered in the gas 
above the cloud before it escapes into space. Because this last
scattering is in the gas, it significantly adds in particular to
the planetary linearly polarized flux $Q$. With increasing
$\lambda$, the amount of light that is scattered in the gas 
above the cloud decreases, and with this, the contribution
to the polarized fluxes and the sensitivity to
the cloud variability decrease.

\citet{Robinson2011} and \citet{Fuji2011}
studied the diurnal variability in Earth's reflected total
flux $F$ and compared results from their numerical 
models to reflected total-flux observations of the Earth taken by the EPOXI mission
(on three separate days in 2008, at $\alpha=57.7^\circ$, 75.1$^\circ$, 
and 76.6$^\circ$, respectively, although \citet{Fuji2011}
did not use the 75.1$^\circ$ data because the moon was partially
in view),
while \citet{2018AJ....156...26J} analyzed reflected
total-flux data of the DSCOVR mission (taken only at $\alpha=0^\circ$).
\citet{Munoz2015Towards} used his radiative transfer code to simulate
2005 MESSENGER data (reflected total fluxes) between $\alpha=98^\circ$ 
to 107$^\circ$.
While quantitative comparisons between our computed reflected total 
fluxes and the measured data and computed fluxes 
presented in these papers is difficult because 
of differences in cloud coverage and orientation of the model
planet (when limited phase angles are covered in the observations),
and because different definitions were used, for instance 
(e.g., fluxes are shown as normalized to a 24h time average).
We can, however, perform a qualitative comparison (only for the 
total fluxes because polarized fluxes of the Earth as a whole,
have not been measured from space). 

In general, \citet{Robinson2011} and \citet{Fuji2011} reported a similar
increase in the peak-to-peak variability of the reflected $F$ data
with increasing $\lambda$ as we, and the wavelength dependence
of $F$ is similar (across the phase-angle range of the EPOXI data);
the Earth is darkest around 650~nm \citep{Fuji2011}
or 670~nm (our simulations).
\citet{2018AJ....156...26J} showed a stronger peak-to-peak variability
in $F$ at 780~nm than at 380~nm at the 0$^\circ$ phase angle
of the DSCOVR data (the spectral solar flux
was included in the reflected flux data in their paper).
Our diurnal curves as shown in Fig.~\ref{fig:PhaseCurves} 
(diurnal curves near full-phase)
agree (qualitatively) well with DSCOVR observations, 
with strong enhancements in $F$ when large convective clouds 
near the maritime continents and 
the Atlantic side of the American continent are in view
\citep[see][]{2018AJ....156...26J}.
The reflected MESSENGER fluxes reported by \citet{Munoz2015Towards}
show the same dependence of $F$ on $\lambda$, but a less obvious
stronger peak-to-peak variability with increasing wavelength. 
This might be explained by the smaller wavelength range 
that is covered by the MESSENGER data, as only 480~nm,
560~nm, and 630~nm are shown by \citet{Munoz2015Towards}. 

\subsection{Color changes}
\label{Sect_Results_Color_Changes}

\begin{figure*}[ht!]
\centering
\includegraphics[width=0.94\textwidth]{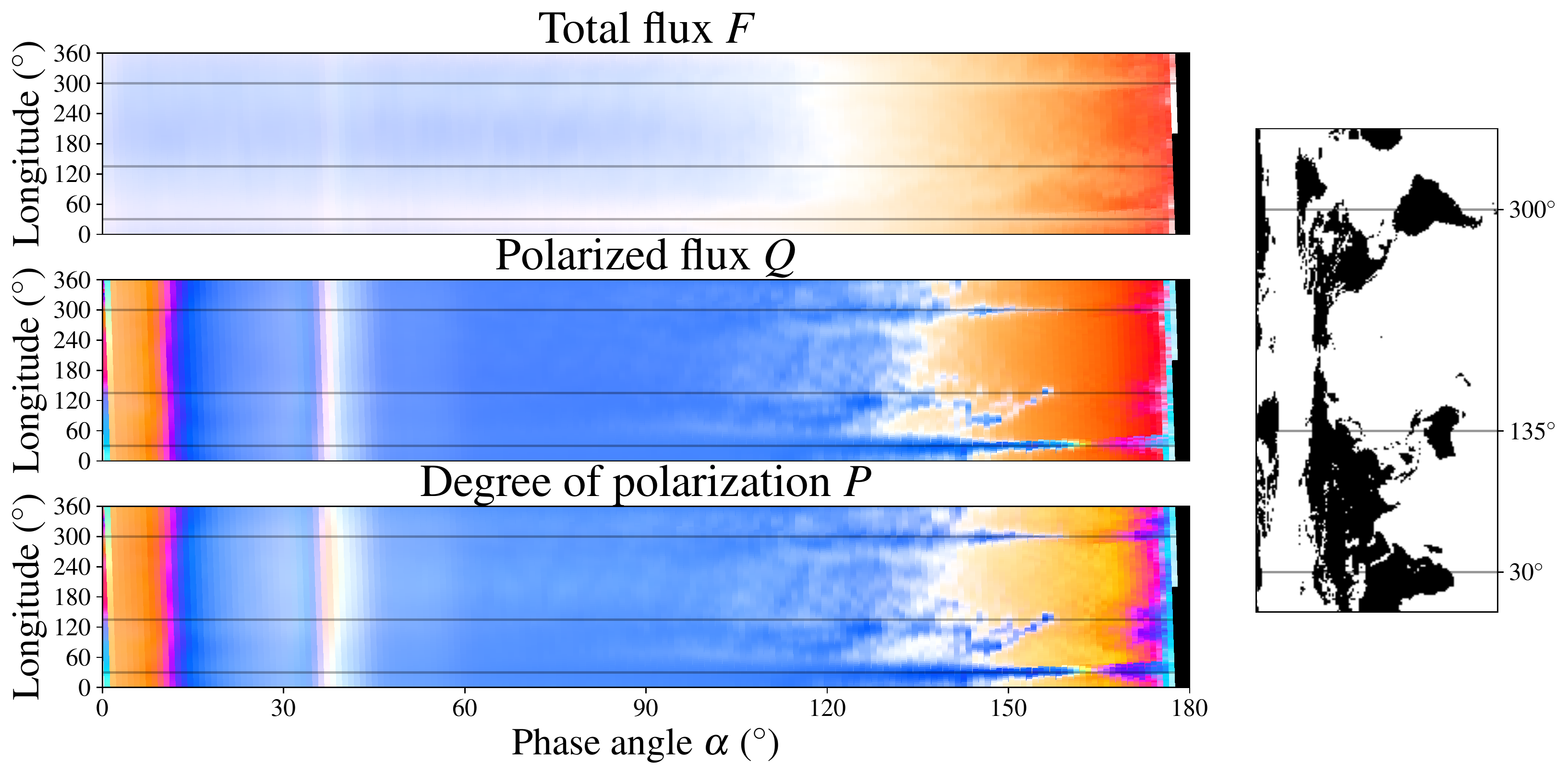}
\caption{The RGB colors of $F$ (top), $Q$ (middle), and $P$ (bottom) of
         the cloudy model planet as functions of phase angle $\alpha$ 
         and the longitude in the middle of the longitude range that 
         is visible to the observer (0$^\circ$ longitude corresponds 
         with the Greenwich meridian).
         These color plots were constructed from the $F$, $Q$, and
         $P$ phase curves at 443, 550 and 670~nm as presented in 
         Fig.~\ref{fig:PhaseCurves}. The horizontal black lines 
         indicate the respective longitudes on the geographical map on 
         the right.}
\label{fig:Lon_phase_Color}
\end{figure*}

In Fig.~\ref{fig:Lon_phase_Color} we show the colors of our cloudy model
planet (with an average cloud fraction of about 66\%) 
in $F$, $Q$, and $P$ as functions of the geographic longitude
in the middle of the longitude range on the illuminated and visible 
part of the planetary disk, 
and phase angle $\alpha$ (these are functions of time 
themselves, i.e.,
with increasing longitude, $\alpha$ also slightly increases),
computed using the curves at 443, 550, 
and 670~nm in Fig.~\ref{fig:PhaseCurves}. 
With increasing $\alpha$, the effect of the local patterns on the
planet starts to appear, as the illuminated and visible part of the 
planetary disk decreases.

Figure~\ref{fig:Lon_phase_Color} shows how the planet changes color in 
$F$ from blue through white to red with increasing $\alpha$. The planet
appears white when the phase curves of $F$ at $\lambda=443$ nm, 550~nm 
and 670~nm intersect 
(cf.\ Fig.~\ref{fig:PhaseCurves}), which is near $\alpha = 120^\circ$,
in Fig.~\ref{fig:Lon_phase_Color}.
The faint white vertical stripe near $\alpha=40^\circ$ is due 
to the primary rainbow. This rainbow feature is much stronger in
$Q$ and $P$. Flux $F$ shows a few whitish, horizontal 
streaks that are related to the presence of continents around
those longitudes, in particular, Africa (around 30$^\circ$) and 
the Americas (around 300$^\circ$). At smaller phase angles,
the brown and green color of the arid and vegetated regions 
comprising the largest regions of the continents 
(see Fig.~\ref{fig:RGBdisks}) 
compensate for the blueness of the atmosphere, while at the
larger phase angles, the continents limit the reflection by the 
ocean that tends to color a planet red at these phase angles
\citep[]{TreesandStam2019}.

The color changes in polarized flux $Q$ and degree of 
polarization $P$ appear to be related to the visibility of an
ocean surface depending on which part of the planet
is visible to the observer. \citet{TreesandStam2019} have
derived that for ocean planets, that is, planets without continents,
the color change phase angle for $F$ decreases with 
increasing cloud coverage fraction $f_{\rm c}$, 
while it increases with increasing $f_{\rm c}$ for $Q$ and $P$.
In Fig.~\ref{fig:Lon_phase_Color}, $F$ changes color in a ~5-10$^\circ$
wide phase-angle range around 120$^\circ$, which on an ocean planet
would be indicative for $f_{\rm c}$ ranging from  
0.25 to 0.75 \citep{TreesandStam2019}. However, on these
model planets, the continental surface coverage and the surface 
albedo of the continents also affects the color of $F$.
According to \citet{TreesandStam2019}, for $Q$ and $P$ of ocean
planets, the phase angle where the color changes increases with
increasing $f_{\rm c}$, and Fig.~\ref{fig:Lon_phase_Color}
shows that adding continents (partly) below the clouds appears to
have a similar effect: for longitudes with more continental
coverage, the color-changing phase angle is larger than for 
longitudes with less continental coverage (assuming that 
the clouds are more or less evenly distributed across the planet).

A tentative color crossing in polarized Earth-shine observations 
has recently been presented by \citet{Sterzik2018}.
These measurements were obtained with ESO's Very Large Telescope (VLT).
Given the telescope location in Chile, either the Pacific Ocean
or South America and the Atlantic Ocean would have been oriented
toward the Moon, depending on the time of the observations.
The Pacific Ocean observations cover the widest phase-angle range
(from about 50$^\circ$ to 135$^\circ$) and show a tentative 
color crossing near 135$^\circ$.
This phase angle is similar to what is shown in 
Fig.~\ref{fig:Lon_phase_Color}.
The Atlantic Ocean observations cover phase angles from 
about 65$^\circ$ to 116$^\circ$, which is not far enough to actually
show a color-crossing phase angle, but extrapolation by eye
suggests a crossing phase-angle between 125$^\circ$ and 130$^\circ$,
which would be smaller than shown in Fig.~\ref{fig:Lon_phase_Color}. 
This might be due to a different orientation of the Earth than
used in our simulations or to different cloud properties.
The precise relation between
the color-changing phase angle, the surface coverage, and the (variable)
cloud coverage and its applicability in characterizing the distribution
of continents and ocean on exoplanets will be subject of a later study.

\section{Summary and discussion}
\label{Sect_Discussion}

We have presented numerically computed total and polarized fluxes of
starlight that is reflected by an Earth-like model planet as functions 
of time, the planetary phase angle, and for five different wavelengths 
in the continuum: 350, 443, 550, 670, and 865~nm.
These fluxes and the derived degree of polarization can play a crucial
role as test data to help the design and development of astronomical 
instrumentation and observational strategies for the direct detection 
of small Earth-like exoplanets, especially because such data (across
time, phase angles, and wavelength) are unavailable from Earth-observation
missions.

Estimates show that with a dedicated technology maturation path, 
strategic exo-Earth characterization missions might be possible around
2030-2035 \citep[][]{Crill2017}. This estimate is based on an extensive
study of current technology areas that must be advanced to image
exo-Earths, albeit including only brightness (total flux) measurement
techniques. Similarly, \citet{2018Feng} estimated the science return 
of space observatories such as HabEx and LUVOIR for different instrument
designs with an inverse-modeling framework. They also considered only total
fluxes. 

With our results, the advantages of including spectropolarimetry can be
studied on the instrument and mission level. In particular, our results can 
be used to investigate the effect of the polarimetric accuracy that can 
be achieved by different polarimetric techniques in an instrument and 
by different telescope designs, especially the effects of various
advanced adaptive optics techniques, and different types of coronagraphs
to reduce the background stellar flux and to optimize the inner working
angle. By integrating our results over time, effects of observational
integration times can be investigated.

Specific phase-angle-dependent features that would be worth detecting
appear to be the primary rainbow as a tool for detecting liquid water clouds
\citep[as earlier identified, e.g., by][]{karalidi2012looking,bailey2007rainbows}, 
which is a stronger signal in $Q$ and $P$ than in $F$, the color-crossing
phase angle in $Q$ or $P$ as a tool for detecting liquid surfaces 
\citep[for details, see][]{TreesandStam2019}, and the regular variations
due to the daily rotation of the planet and the variability on those
variations due to the changing cloud cover. When the 
orientation of the planetary light equator with respect to the 
instrumental optical plane is not known, because the orientation of the
planetary orbit is not known, the distribution of linearly polarized flux
over $Q$ and $U$ is generally different than in our results. The 
very low values of $|U|$ as compared to $|Q|$ when the instrumental
reference plane is parallel to the light equator, even for a horizontally
inhomogeneous planet like the Earth, could be used to determine the
orientation of the light equator, and with this, of the orbit.

Because the diurnal variability of the signals appears to be little
dependent on the wavelength, except for $|Q|$ at short wavelengths,
broadband observations would help increase the signal-to-noise ratio of
observations
\citep[see][for a similar discussion on the observability of the rainbow feature]{bailey2007rainbows}. 
At short wavelengths ($\lambda$ = 350~nm in our results), the variability
of $|Q|$ strongly depends on the cloud properties and very little on 
the surface properties. Measuring in a short and a long wavelength band
would therefore allow for uncoupling cloud and surface reflection. The
color-crossing $\alpha$ for liquid surface detection is at a relatively
large phase angle, that is, near 145$^\circ$ for an Earth-like cloud 
coverage fraction
\citep[the larger the coverage fraction, the larger the color-crossing phase angle, see][]{TreesandStam2019}, and $F$, $|Q|$ and $P$ are 
relatively small at these phase angles. In addition, the color of the
planet should be determined, and observations through more than one
spectral band should therefore be performed. However, the color-crossing can be 
detected across a range of phase angles of 10$^\circ$-20$^\circ$~wide,
which would allow for long integration times, see 
Fig.~\ref{fig:Lon_phase_Color}. 

Our polarized flux results are valuable test data not only for
spectropolarimeters, instruments designed to measure polarization, but
also for spectrometers, instruments that might not have been designed to
measure polarization, but that are often polarization sensitive due to 
the use of optical elements such as mirrors and gratings. Without a 
proper validation of the design of a spectrometer for incident light 
that is polarized, such as starlight that is reflected by an exoplanet,
measurement errors arise that are hard to account for. When test data are used
in combination with instrument simulators, the errors for a given design
can be estimated and the design can be adapted to minimize them.

Because the physical parameters of our model Earth are known, our 
computed fluxes and the degree of polarization across time, phase angle, 
and wavelength can also be used to develop and/or test data analysis
algorithms to be used on (future) measurements of light reflected by
exoplanets, such as autocorrelation methods of the Fourier analysis to 
derive the planetary rotational period and/or the presence of continents
\citep[see][]{2015A&A...579A..21V,Oakley2009,Palle2008}. To simulate
actual measurements, our total and polarized fluxes would have to be
scaled to those of the target exoplanet and its star, according to
Eq.~\ref{eq:reflected_fluxes}, noise sources (photon noise and 
background stellar light) and instrumental effects (e.g.\ derived from 
a preliminary design) could be added, by scaling the computed fluxes
and/or integrating across instrument response functions, effects of
different integration times and/or spectral resolutions can be studied.
Our computed total and polarized fluxes could also be combined with 
total and polarized fluxes of light of the parent star that is scattered
by exo-zodiacal dust
\citep[see, e.g.,][]{1995A&A...304..602R} to investigate the effect
of such local background sources on the detectability of the 
planetary signals.

For our model planet, we used Earth-observation data covering the
period from January 1, 2011, to December 31, 2011, to describe the
surface coverage and local albedos, the cloud coverage fraction, 
optical thickness, top pressure, and the effective radius of the cloud
particles. The surface types that we took into account are oceans (with
waves), vegetation (deciduous forest, grass, and steppe), arid regions 
(desert and shrublands), and ice or snow. The reflection by the ocean and
arid regions fully includes linear (and circular) polarization, while 
that by the vegetation only includes linear polarization. Ice and snow
are described by a Lambertian, that is, isotropic and unpolarized. 

For the numerical results presented here, we did not include scattering 
and absorption by aerosol particles (small, suspended particles in the
atmosphere): although numerically they are straightforward to include,
their variations in space and time would require the addition of many
atmosphere-surface models and hence severely increase the computing times.
\citet{2011ApJ...739...55Z} and \citet{zugger2010light} investigated 
the effects of spherical maritime (sea-salt) aerosol particles (which are
likely nonspherical \citep[see][]{BookMishchenko2000}) 
on the polarized reflection of ocean planets and concluded that depending
on the visibility, these aerosols could significantly depolarize the 
reflected fluxes, although a comparison between the Earth-shine 
observations and model computations without aerosol presented by 
\citet{Sterzik2018} strongly suggests that aerosols in general have a
minor effect on the polarized appearance of Earth, especially when
compared to the effects of clouds (which \citet{2011ApJ...739...55Z} 
and \citet{zugger2010light} did not combine with the maritime aerosol).
Our model atmospheres do not include (nonspherical) ice cloud
particles either; all cloud particles are liquid water droplets. Scattering by
nonspherical ice cloud particles was included by 
\citet{karalidi2012looking}, who used both liquid water clouds and ice
clouds in their model atmospheres. The ice- and liquid-cloud optical
thicknesses were derived from MODIS data. They adopted imperfectly shaped 
hexagonal ice-cloud particles \citep[][]{1998JQSRT..60..301H}, and 
concluded among others, that the presence of ice clouds does not limit 
the visibility of the rainbow (which is due to spherical particles alone)
in the polarization. \citet{2017A&A...605A...2E} used both types of 
cloud particles and relatively high ice-cloud optical thicknesses
in their analysis of the polarized Earth-shine data published by 
\citet{2012Natur.483...64S}.

Our numerical computations do not include the absorption of light 
by atmospheric gases either, such as water and oxygen. This absorption can be
included and will lead to features with high spectral resolution polarization
across the absorption bands 
\citep[see, e.g.,][]{2017ApJ...842...41F,stam2008spectropolarimetric}, 
so that our fluxes should be considered to be representative of the continuum.
Absorption by ozone is also neglected. This would be important to be included 
when wavelengths below 350~nm are of interest. The wavelength of 
550~nm falls within a broad diffuse ozone absorption band (the Chappuis 
band) that would lower the reflected total and polarized
fluxes, especially in the presence of clouds, but it would hardly 
change the degree of polarization \citep[see, e.g.,][]{stam2008spectropolarimetric}.
We also ignored rotational Raman scattering, which is an inelastic scattering 
process by gaseous molecules, which would leave imprints of the stellar
Fraunhofer lines in the total and polarized fluxes and in the degree 
of polarization as functions of the wavelength. In Earth observations,
this phenomenon is usually referred to as the Ring effect after one 
of the discoverers \citep[][]{1962Natur.193..762G}.
Numerical simulations of rotational Raman scattering in atmospheres
of gas giants (in total fluxes only) has been presented by  
\citet{2016ApJ...832...30O}. Rotational Raman scattering could
be added to our numerical algorithm, but would require significant amounts
of computing time.

Some improvements of our descriptions of the surface reflection are
the reflection by the surface below shallow waters, altitude variations
on the surface, such as mountains, 
regions of the ocean covered by sea ice,
and seasonal variations in the surface reflection. 
Regarding the latter, in particular, seasonal albedo changes due to
the migration of sea ice over the ocean, for instance,
changes in the colors of vegetation,
in the bloom of algae, etc.\
could be implemented. Polarization measurements of locally scattered 
and/or reflected light, for example, by instruments
on board satellites in low Earth orbits, would be very useful 
for the extension and/or improvement of the currently missing or
insufficient numerical descriptions 
of scattered or reflected light, as described above.

While validation of computed disk-integrated total and/or
polarized fluxes reflected by a model Earth can be made somewhat using 
(polarized) Earth-shine measurements \citep[see][]{2012Natur.483...64S}, 
this method has several limitations,
in particular, the unknown influence of the reflection of Earth-light
by the lunar surface. While we can in principle measure how
the surface (at different locations on the Moon) reflects incident
(unpolarized) sunlight for various local solar zenith angles and 
various local viewing angles, we do not know how the surface 
reflects polarized light, especially not as a function of wavelength
and covering a range of illumination and viewing geometries. 
\citet{Sterzik2018} presented a method to correct their polarized
Earth-shine measurements for the lunar depolarization effect,
but the uncertainty in this method appears to be too large to
validate parameter choices for the model planets.
The ideal validation data would be spectropolarimetric measurements 
of the Earth taken from a distance. 
A compact instrument that would provide such data is 
the Lunar Observatory for Unresolved Polarimetry of Earth (LOUPE)
\citep{LOUPE2016,Karalidi2012LOUPE}, which aims to 
measure the total flux and the degree and direction of linear
polarization from about 400 to 800~nm as the payload of a lunar 
lander or a geostationary satellite, for example (the latter would, however, 
not allow monitoring signal variations because the planet rotates daily).
The spectropolarimeter LSDpol also aims to measure circular polarization
across a similar spectral range 350-900 nm \citep{Snik2019}.

The computed total and polarized fluxes are available through 
the database researchdata.4tu.nl. They are archived under 
'Colors of an Earth-like exoplanet' with DOI
10.4121/uuid:caa03e1a-0f6a-43e5-b67a-fa6f21829b8a 

\begin{acknowledgements}
The MODIS Land Cover Type Data set (2011) is maintained by the NASA 
EOSDIS Land Processes Distributed Active Archive Center (LP DAAC) at 
the USGS/Earth Resources Observation and Science (EROS) Center.
The Aqua/MODIS and Terra/MODIS Cloud Top Pressure, Cloud Particle
Effective Radius, Cloud Optical Thickness and Cloud Fraction Daily L3
Global 1 Deg. CMG datasets were acquired from the Level-1 and Atmosphere
Archive \& Distribution System (LAADS) Distributed Active Archive Center
(DAAC), from the Goddard Space Flight Center in Greenbelt, Maryland, USA
(https://ladsweb.nascom.nasa.gov/). The CCMP Version-2.0 vector 
wind analyses are produced by Remote Sensing Systems, with data available
at www.remss.com. We acknowledge the use of the SCIAMACHY surface LER 
database provided by the Royal Netherlands Meteorological Institute 
(KNMI). L.\ Rossi acknowledges funding through the Planetary and 
Exoplanetary Science (PEPSci) Programme of the Netherlands 
Organisation for Scientific Research (NWO).
\end{acknowledgements}


\end{document}